\documentclass[
 reprint,
 amsmath,amssymb,
 aps,pra,superscriptaddress,nobibnotes
]{revtex4-2}
\usepackage{amssymb}
\usepackage{amsmath}
\usepackage{cancel}
\usepackage{color}
\usepackage{braket}
\usepackage{graphicx}
\usepackage{epstopdf}
\usepackage{verbatim}
\usepackage{siunitx}
\usepackage{soul}
\usepackage{xr}
\usepackage[toc,page]{appendix}
\usepackage[section]{placeins}
\usepackage[utf8]{inputenc}
\usepackage[colorlinks]{hyperref}

\DeclareUnicodeCharacter{2009}{\,}

\newcommand{\order}{\mathcal{O}}

\begin{document}

\title{Thermodynamic Limits of Quantum Search}

\author{R. Riedinger}
\email{ralf.riedinger@uni-hamburg.de}
\affiliation{Institut für Quantenphysik und Zentrum für Optische Quantentechnologien, Universit\"{a}t Hamburg, 22761 Hamburg, Germany}
\affiliation{The Hamburg Centre for Ultrafast Imaging, 22761 Hamburg, Germany}
\date{March 13, 2026}

\begin{abstract}
Modern cryptography relies on keyed symmetric ciphers to ensure the secrecy and authenticity of high bandwidth data transfer. 
While the advent of quantum computers poses a challenge for public key cryptography, unbroken ciphers are considered safe against quantum attacks if their key is sufficiently long. 
However, concrete bounds on the required key length thus far remain elusive: 
Despite the well known asymptotic complexity of Grover's quantum search, the optimal algorithm to recover a secret key, no implementation-agnostic tight bounds were established. 
Here, we discuss the quantum thermodynamic limits of generic search algorithms, and find a work-runtime trade-off for autonomous computers with a fundamental lower bound. By devising an application-specific quantum protocol, which outperforms circuit and adiabatic implementations, and saturates this bound, we demonstrate that it is tight. 
Applying this limit, we find that a secret key of 831 bit length cannot be reconstructed deterministically in an expanding, dark-energy-dominated universe until star formation is expected to cease. 
Implications for post quantum cryptography, and quantum key distribution are discussed.

\end{abstract}

\maketitle
\section{Introduction}
Conventional computers use irreversible operations, like the NAND gate, which has two input bits, but only one output. 
The information lost in the process increases the entropy of the system, respectively, generates heat. Landauer realized that this poses a fundamental limit of irreversible computation, which can be expressed as the work required to keep the system at a constant temperature \cite{landauer_irreversibility_1961}. 

While current systems are far from reaching the Landauer limit, it still serves as a hard bound for the capacity of (irreversible) computers, for example for exhaustive search of cryptographic keys. This allows for defining concrete security levels for unbroken symmetric ciphers, assuming only the work available to an adversary, and the irreversibility of conventional logic gates. 

Quantum computation, in contrast, is natively a unitary operation and thus reversible. While error correction can add an irreversible component, there is no fundamental lower limit to it, and consequently, and thus far, no strict energy bounds have been established. It has even been argued, that energy considerations by themselves are insufficient to establish computational bounds \cite{Jordan2017}. 

Here, we demonstrate that a fundamental thermodynamic energy bound exists for reversible quantum search, too. The bound represents a new type of quantum speed limit and is independent of the specific implementation of Grover's algorithm  \cite{grover_fast_1996, grover_quantum_1997}, merely assuming a suitable oracle is available. It can be adapted to other algorithms, which is, however, left for future work.

We apply our quantum search bound to establish how long a cryptographic key need to be, for it to be considered quantum resistant. 
Notably, we find that doubling the key length in insufficient, to provide a comparable level of security in the post quantum age. 
This common believe \cite{chen_report_2016, jaeger_quantum_2021, jaques_implementing_2020} is based on the fact that Grover's algorithm, which is asymptotically optimal \cite{zalka_grovers_1999}, can find an $n$-bit key with $\order(2^{n/2})$ gates, in contrast to the $\order(2^{n})$ operations required for classical exhaustive search. Implicitly, this assumes future quantum computers can achieve gate depths similar to current conventional computers.
As unitary gates do not consume work, any assumption on gate depth is technological in nature, and thus subject to speculation. For example, it can and has been argued, that  
quantum algorithms are hard to parallelize and hence shorter keys may be sufficient \cite{perlner_thermodynamic_2017, fluhrer_scott_reassessing_2017}. 
Estimates of key size requirements are often based on the specific implementations of computer architectures and algorithms \cite{boyar_small_2012, grassl_applying_2016, boyar_small_2019, langenberg_reducing_2020, jaeger_quantum_2021, alagic_status_2022}.
It is, however, notoriously hard to predict the trajectory of a new technology, where efficient application-specific implementations can tremendously change underlying assumptions of what is realistic \cite{bhattacharya_implementation_2002, molmer_efficient_2011, langenberg_reducing_2020}.

The fundamental thermodynamic limit established here allows for determining concrete recommendations for quantum resistant key lengths independent of techonological implementation: The security level is determined by the work and time an classical or quantum adversary is able to invest into key recovery, as well as an acceptable success probability under these conditions. Based on these assumptions, we find that classically secure key lengths need to be approximately tripled to provide the same thermodynamically limited security against a quantum attacker. 

Further, we estimate the number of bits, beyond which extending a key provides diminishing added security from a thermodynamic point of view, by accounting for all energy inside the cosmic event horizon. This maximum useful key length, estimated to be $n_\text{max}\sim 829$~bits, provides an important reference for any application attempting to provide strong privacy, such as post quantum cryptography or quantum key distribution (QKD). 

First, we review existing thermodynamic limits for classical and quantum algorithms for exhaustive search, and then derive an system-agnostic bound for it. Using an example implementation, we demonstrate that the bound is tight. Finally, recommendations for quantum resistant key-lengths with security comparable to conventional cryptography standards are discussed.

\section{Limits of classical search}
For a classical, irreversible machine, two fundamental limits of computation are relevant: (1) Landauer's principle \cite{landauer_irreversibility_1961} states that whenever a bit is reset, the entropy of a reservoir with temperature $T$ increases, requiring an energy of $E_L\geq k_BT\ln 2$. (2) The Margolus-Levitin theorem \cite{margolus_maximum_1998, okuyama_quantum_2018, shanahan_quantum_2018} states that a system when changing between orthogonal states within time $\delta t$, needs an average energy of $E_{ML}\geq h/4\delta t$. 
Other quantum speed limits \cite{deffner_quantum_2017}, like the Mandelstam-Tamm bound \cite{mandelstam_uncertainty_1945}, can further bound the machine, but do not impose stricter energy limits (see \ref{ap:classic}). When imposing size limits, Bremmerman-Bekenstein bounds can impose further entropy-based restrictions, which we neglect here \cite{Deffner2020}. 
Hence, the minimum energy such a machine requires to recover a $n$-bit key from an $n$-bit cipher text with probability $P_s$ in time $t$ is
\begin{equation}\label{eq:classical_limit}
    E_{c}\geq 2^n P_s \left( k_B T \ln{2} + \frac{h}{4t}\right) + 2n k_BT\ln 2,
\end{equation}
when running at constant speed ($1/\delta t = t/2^n P_s$), see \ref{ap:classic} for details. For example, an irreversible classical machine, aiming to reconstruct a 256-bit key with a probability $P_s\geq 10^{-10}$ would require the entire radiative energy the sun emits during the remainder of its lifetime ($t\sim 5$Ga), using the cosmic microwave background ($T_\textrm{CMB}=2.7$K) as cold bath\cite{Fixsen_2009}, see also table \ref{tab:quantum_key_lengths}.

An adversary with access to a quantum computer can employ Grover's algorithm \cite{grover_fast_1996}, which yields a result after $\order(\sqrt{2^n})$ iterations.
More importantly, however, they may hardly be restricted by above limits: As quantum search is a unitary evolution, merely qubit initialization and readout are necessarily irreversible, and thus subject to the Landauer, or Bremmerman-Bekenstein bounds \cite{Deffner2020}.
Furthermore, the system transfers to a (near-)orthogonal state only once during an adiabatic computation (initial state to solution state), such that quantum speed limits do not provide a strict bound either.
In summary, the energy bounds for Grover's search based on the Landauer- and Margolus-Levitin-limit $E_\textrm{q,LML}\geq 2\left( n k_B T \ln{2} + \frac{h}{4t}\right)$ are exponentially lower than for classical search. 

The adiabatic quantum search \cite{roland_quantum_2002}, suggests an energy bound $E_A\geq \order(2^{n/2}/t)$ to ensure an upper bound to the computational error, albeit resonances with substantially lower error probability can be found numerically \cite{rezakhani_accuracy_2010}. Hence, this cannot be used directly to establish a lower energy bound, either.

\section{Limit of dynamic evolution in autonomous machines}
To establish an implementation-agnostic tight lower bound on the resources required by quantum search, we first introduce our notion of an autonomous quantum computer, and derive a quantum speed limit stated as a work requirement for non-orthogonal wavefunction evolution. Then, an implementation agnostic description of Grover's algorithm introduced and optimized with regards to fundamental resource requirements. 

We describe an autonomous quantum computer by the time independent Hamiltonian 
\begin{equation}
    H = H_Q + H_C
\end{equation}
consisting the computational Hamiltonian $H_Q$, governing the dynamics of the computational space, a control Hamiltonian $H_C$, describing all necessary external control systems, for example lasers used to drive gates, or adiabatically varying interactions. 
For the sake of simplicity, we assume that the autonomous computer is a sufficiently compact and closed system for the relevant time scale. 
This ensures no energy requirements can be hidden in the control system or a bath, which can lead to increased speed limits in open systems \cite{deffner_quantum_2013, cimmarusti_environment-assisted_2015}. 

In such a closed system, $\left[H,H^m\right]=0$ for all integers $m$, meaning energy, and its all moments are conserved. 
A macroscopic energy source, like a barrel of oil or a battery, which can be employed to power the machine, has a low relative energy uncertainty $\Delta E/E\sim1/\sqrt{N_\textrm{DOF}}$, with $N_\textrm{DOF}$ degrees of freedom, assuming it is in a thermal state, with most of its energy stored as potential energy (see SI). This motivates the investigation of the thermodynamic work $W$ required to operate the computer, which is defined with no higher momenta, i.e.\ $\Delta W\approx 0$. 
Formally, we consider the autonomous computer to start at temperature $T=0$, and a charged battery as part of the control system, providing work $W$. 
This computer is tasked with preparing a state $\ket{\psi}$.
Its highest energy component is $\ket{\omega_M}$, the eigenstate of $H$ with the largest energy $\hbar \omega_M$ and non-zero overlap $\langle\psi\mid \omega_M\rangle\neq0$. 
The work required to prepare this state is
\begin{equation} \label{eq:work}
   W_\psi \geq \sqrt[m]{\langle H^m \rangle}  \geq \left|\langle\psi\mid \omega_M\rangle\right|^{\frac{2}{m}} \hbar \omega_{M}  \xrightarrow{m\rightarrow\infty} \hbar \omega_{M},
\end{equation}
due to the conservation of the $m^\textrm{th}$ energy momentum.

A lower bound on the highest energy component $\hbar \omega_{M}$ can for example be obtained from the highest frequency component of the expectation value $\langle O\rangle$ of any time-independent observable in the system, such that inequality \eqref{eq:work} can be interpreted as a new type of quantum speed limit. In an autonomous machine, this explicitly includes all control fields. For example, Grover's gate-based quantum search requires $W_Gt\geq \hbar \sqrt{2^n} \pi-\epsilon$ due to the minimum bandwidth of the control fields ($\epsilon$: small offset, see 
\ref{ap:gates}).
Applied to adiabatic computing, 
we find that in addition to the ground state energy \cite{roland_quantum_2002}, the bandwidth of the control fields imposes energy requirements (see \ref{ap:adiabatic}). 
The inequality \eqref{eq:work} differs from other speed limits, as it sets a bound for the work required to create a state. This results in a stricter bound, circumventing situational limits, as is the case for Mandelstam-Tamm and Margolus-Levitin bounds \cite{ness_observing_2021}, as well as increased limits for open non-Markovian systems\cite{deffner_quantum_2013, cimmarusti_environment-assisted_2015}. 

The bound can potentially be slightly softened when allowing for a thermal machine with a finite uncertainty in the energy reservoir. However, when considering a non-vanishing observable frequency components, this reduces the energy $E_\psi$ required for state preparation only by $1-E_\psi/W_\psi = \order\left(N_\textrm{DOF}^{-1/2}\right)$, a negligible difference for macroscopic systems with a large number of degrees of freedom $N_\textrm{DOF}$.

\section{A tight lower bound for quantum search}
After having established that gate-based and adiabatic quantum search follow a similar scaling, we want to establish a tight, implementation-agnostic lower bound, to determine quantum-resistant key sizes. 
The general autonomous quantum search algorithm should find a solution state $\left|s\right\rangle_Q \in |\{0,1\}^n\rangle_Q$ in the $n$-qubit computational space representing the secret key, given the oracle operator $|s\rangle_Q\langle s|_Q$. We require the optimized algorithm to achieve this in the minimum time possible for a fixed input work $W$, independent of the specific solution. 

These conditions demand the system to be initialized in an equal superposition of all possible solutions 
$|i\rangle_Q \propto \sum_{k\in\{0,1\}^n} |k\rangle_Q$ 
and restricts the computational Hamiltonian configurations to (see \ref{ap:eq45})
\begin{equation}
    H_Q=\hbar\omega_i \ket{i}\bra{i} + \hbar\omega_s \ket{s}\bra{s} + H_\perp. \label{eq:gen_ham}
\end{equation}
Here, we dropped the index $Q$ indicating the computational space, and set all relative phases to 0, without loss of generality. $H_\perp$ only affects states orthogonal to the space spanned by $\ket{i}$ and $\ket{s}$ and contains arbitrary offsets. As the computational state 
$|\psi\rangle$ 
stays within the $\ket{i}$-$\ket{s}$-space, $H_\perp$ can be neglected. 

In this formalism, adiabatic quantum search is obtained by slowly changing  $(\hbar\omega_i, \hbar\omega_s)=(E,0)\rightarrow(0,E)$ with a characteristic energy $E$. Grover's original bandwidth-unlimited implementation is recovered by switching $\pi\sqrt{2^n}/4$ times between $(\hbar\omega_i, \hbar\omega_s)=(0,E)$ and $(E,0)$ for $\delta t=\hbar\pi/2E$ each. 

To find the optimal configuration, we aim to increase the overlap with the solution $|\langle s\mid \psi\rangle|$ as fast as possible under the finite work constraint. 
Using \eqref{eq:gen_ham} and Schr\"odinger's equation, we find (see \ref{ap:eq45})
\begin{equation}\label{eq:rateofchange}
    \frac{d}{dt} |\langle s\mid \psi\rangle| =  \frac{\omega+\delta\omega}{\sqrt{2^{n}}} \sin(\alpha_{ab}) |\langle i\mid \psi\rangle| 
\end{equation}
with $2\omega=\omega_i+\omega_s$, and $2\delta\omega=\omega_i-\omega_s$. The relative phase $\alpha_{ab}=\arg(\langle i\mid \psi\rangle\langle \psi\mid s\rangle)$ between the computaional state's overlap with the initial and solution state can provide guidance on the effectiveness of quantum search implementations:

As the coupling $\left\langle i \right|H_Q\left| s \right\rangle$ in \eqref{eq:gen_ham} is real-valued, the eigenstates of $H_Q$ have $\alpha_{ab}=0$ for all $\omega_i,\omega_s$. Adiabatic search protocols, require the computational state to stay close to an eigenstate, resulting in  $\sin(\alpha_{ab})\ll1$, and thus intrinsically slow information gain. 

The detuning determines the phase velocity
\begin{equation}
    \label{eq:phasevelocity}
\frac{d}{dt}\alpha_{ab}\approx-2\delta\omega
\end{equation}
under most circumstances ($|\delta\omega/\omega|, |\langle s\mid \psi\rangle\langle i\mid \psi\rangle|\gg1/\sqrt{2^n}$). Substantial information gain over a time interval $\Delta t$ can be maximized by keeping the time average $\langle\sin(\alpha_{ab})\rangle_{\Delta t}$ large. With \eqref{eq:phasevelocity}, we find for the majority of the quantum search, this requires keeping the detuning $\delta \omega$ small. 
This can for example be achieved by resonantly modulating $\omega_i(t')$ and $\omega_s(t')$ for $0\leq t'\leq t$, as done in Grover's gate-based search, see also \ref{ap:eq45}.

We find that during the entire algorithm, the information gain is maximized locally for $\delta \omega\Delta t\leq 16/\omega\Delta t$, see \ref{ap:eq45}. This means a quantum search which runs long enough to establish a non-negligible overlap with the solution state $|\langle s\mid \psi\rangle|$, requires the detuning to be minimal. The potential improvement in the information gain, however, rapidly disappears for longer $\Delta t$, with the relative gain scaling with $(4/\omega\Delta t)^2$. Hence, the opportunity cost represented by the work required to generate the control frequency $\omega_\text{ctrl}\propto 1/\Delta t$, always outweighs the advantage, such that we can limit the detuning to $\delta \omega \leq 16/\omega t^2$ with the total computational time $t$.

Recalling \eqref{eq:work}, and $\delta\omega, \omega_\text{ctrl}\approx 0 $, the average frequency is bound by $\omega\leq W/\hbar (1+2^{-n/2})$, and the average information gain  \eqref{eq:rateofchange} is bound by

\begin{equation}\label{eq:rateofchange3}
    \left\langle\frac{d}{dt} |\langle s\mid \psi\rangle|\right\rangle_{\Delta t} \leq  b\frac{W}{\sqrt{2^{n}}\hbar } |\langle i\mid \psi\rangle| 
\end{equation}
where the prefactor $b\approx 1-\frac{1}{\sqrt{2^n3}}$ is close to unity (see \ref{ap:fundamental_limit}). 
Integrating eq. \eqref{eq:rateofchange3}, we find the lower bound on the required work to find the solution $|s\rangle$ with probability $P_s$ in time $t$ is
\begin{equation}
     W \geq  \sqrt{2^{n}P_s-1}\frac{\hbar}{t} \label{eq:qlimit}.
\end{equation}
Here, we neglected the energy required for initialization and readout (see Appendix \ref{ap:init} for details).

\section{Ballistic Quantum Search}
This limit \eqref{eq:qlimit} can approximately be reached by setting $\delta\omega=0$ for the entire protocol: The system is initialized in $\ket{i}$ and then evolves freely under the constant Hamiltonian $H_Q = \hbar \omega (\ket{i}\bra{i} + \ket{s}\bra{s})$. After a time $t_F=\frac{\pi}{2\omega}\sqrt{2^n}$, it deterministically reaches $\ket{s}$.
As its Hamiltonian is time-independent ($d\delta\omega/dt=0$), no control overhead is fundamentally required ($\omega_\textrm{ctrl}=0$), and the maximum eigenfrequency is minimized. Integrating the ballistic evolution of the state, we find
\begin{equation}\label{eq:ballistic}
    P_S(t) \leq \frac{1}{2^n} + \left(1-\frac{1}{2^n}\right) \sin^2\left(\frac{W}{\sqrt{2^n}+1}\frac t \hbar\right),
\end{equation}
for $0\leq t\leq t_F$, where we set 
$\omega\leq W/(1+2^{-n/2})$, again neglecting readout and initialization.
This reaches the fundamental bound \eqref{eq:qlimit} within $\order(2^{-n/2})$, proving that the bound is tight.
Compared to an optimized gate-based implementation with the same work $W$ available, it can reach the solution $\pi$ times faster.
Effectively, this describes an application specific quantum simulator, tailored to the search problem at hand. As such, any adiabatic quantum computer could be modified to execute the ballistic search, rather than the comparatively slow adiabatic search.  

\section{Quantum resistant key lengths}
Similar to the classical case, the fundamental work requirement for an exhaustive quantum search \eqref{eq:qlimit} can be employed to determine how long a key should be, to be considered quantum-safe. 
In the following, we discuss (1) key lengths that provide a similar security, on a fundamental perspective, as current classical standards, and (2) extrapolate how long a key needs to be, such that it can be considered fundamentally secure. 

In symmetric cryptography, two security levels are often recommended: $n=128$ bit for standard communication, and $n=256$ bit for high security applications. These can be related to physical energy scales based on the Landauer limit \eqref{eq:classical_limit}. This typically requires a somewhat arbitrary choice of power, time-scale, and acceptable probability that a secret key is reconstructed. Of course, neither of them are hard numbers, but the order of magnitude can serve as an indication of the level of security achieved by a certain key length. Here, we compute the key length that provides the same level of security under the same assumptions, but against a quantum adversary. 

For the 128 bit keys, we consider a large-scale data center on earth (environmental temperature $T=\SI{300}{\kelvin}$) with a nominal power consumption of \SI{65}{\mega\watt}, spending 5 years to recover a single key by brute-force search. According \eqref{eq:classical_limit}, the success probability of this attempt is less than 1\% using a classical, irreversible computer, such that it seems unreasonable for anyone to attempt this. However, using a quantum computer, they could be reconstructed deterministically within less than \SI{1}{\nano\second}, with a power supply of \SI{6.5}{\kilo\watt}. While building a quantum computer capable of doing this may be hard, the possibility cannot be disregarded based on power consumption alone. 

It is often suggested, that doubling the key length renders symmetric encryption quantum secure. However, fundamentally, a hypothetical \SI{65}{\mega\watt} quantum data center could potentially reconstruct a 256 bit key in less than \SI{0.1}{\second}. To achieve the same level of security ($P_s\leq1\%$ in $t=\SI{5}{a}$), the key needs to be 394 bit long, i.e.\ approximately tripled in length. For collision resistance of hash functions, for example required in signature algorithms or block chains, we find that an equivalent security, corresponding to NIST PQC security level 2 requires an image space of at least 415 bit (see \ref{ap:bht}). 

As the above scenario is unlikely, but not impossible, 256 bit keys are recommended for classical high security applications. In this case, the entire radiative output of the sun over its residual lifetime limits the success probability of a classical computer to less than 0.03 parts per billion. Using such a Dyson sphere as power source, a quantum computer would still require at least several years to reconstruct a key of twice the length, 512 bit. The same level of security ($P_s\leq3\cdot10^{-11}$ in $t=\SI{5}{\giga a}$), however, requires a key length of 667 bit.

\begin{table}
    \centering
    \begin{tabular}{|c|c|c|c|c|c|}
         \hline
         classical &  &  & Recovery &  & \shortstack{equivalent\\quantum-} \\
         key  & Work  & Time & Probability & Scenario & secure \\
         length & $W$ & $t$ & $P_s$ &  & key length\\\hline
         128 & $10^{16}\si{\joule}$ & \SI{5}{a} & 1\% & \shortstack{\vspace{2pt}\\data\\center} & 394 \\
         256 & $8\cdot 10^{43}\si{\joule}$ & \SI{5}{\giga a} & $3\cdot 10^{-11}$ & \shortstack{\vspace{2pt}\\Dyson\\sphere} & 667\\
         - & $4.6\cdot 10^{69}\si{\joule}$ & \SI{100}{\tera a} & $10^{-12}$ & \shortstack{\vspace{2pt}\\cosmic\\horizon} & 872\\\hline
    \end{tabular}
    \caption{Comparison of key lengths providing similar security agains classical or quantum attackers under different scenarios, corresponding to NIST PQC security categories 1 (128 bit), 5 (256 bit) and the cosmic thermodynamic limit. \vspace{10pt}}
    \label{tab:quantum_key_lengths}
\end{table}

One can further ask when there is no advantage in extending the key further, from a fundamental perspective. For this, we consider the mass-equivalent energy inside the cosmic event horizon, i.e.\ all energy that could somehow be made available to a quantum adversary on earth, and give it time to find a key until star formation ceases. As we are interested in the order of magnitude, we use crude approximations, see \ref{ap:key_lengths} for details. Assuming all baryonic and dark matter within the cosmic event horizon is converted into energy, which is used create a single quantum computer, searching for a single 872 bit key, this cannot succeed with a probability larger than 1 part per trillion. Any keylength $>830$ bit cannot be reconstructed deterministically under these conditions. 
What do we learn from such a purely theoretical Gedankenexperiment? Extending a key beyond 872 bit adds negligible security to a symmetric encryption system, other than patching security issues of the encryption algorithm. 

In particular, key extension mechanisms, such as remote quantum key distribution, do not add security beyond the initial pre-shared key they require, from a thermodynamic standpoint. Unless information theoretically secure one-time-pad encryption is used, the security is limited by the symmetric encryption algorithm. 
Conversely, a sufficiently secure symmetric cipher implies that a pre-shared kilobit key can carry enough entropy for unlimited encrypted communication, aside from side-channel attacks on the hardware. Clearly, key rotation has its merits, nonetheless, for example to limit the vulnerability to physical attacks, or to enhance the security of a cipher. 

\section{Conclusion}
We discussed fundamental thermodynamic limits of quantum search algorithms, and identified a lower bound on the work required to run a search. By constructing an algorithm which saturates this limit, we demonstrated that the bound is tight. This ballistic quantum search can outperform conventional, gate-based Grover algorithm by a constant factor, provided the same thermodynamic work available. We applied this bound to derive quantum secure key lengths, demonstrating that doubling the key length of classical standards is insufficient, from a thermodynamic point of view.
Based on current cosmological models, we further showed that an exhaustive search of more than 830 bit key-space is not deterministically possible in a universe with accelerating expansion. 
This is instructive for the development of quantum secure cryptography, providing a benchmark for future-proof systems. It further highlights that extending a key beyond $\sim$1~kbit does not increase the security in itself.

\begin{acknowledgements}
I thank Pius Gerisch, Manik Dawar and Marcus Huber for discussions. R. R. acknowledges support from the Cluster of Excellence 'Advanced Imaging of Matter' of the Deutsche Forschungsgemeinschaft (DFG) - EXC 2056 - project ID 390715994, 
Bundesministerium für Forschung, Technologie und Raumfahrt (BMFTR) via project QuantumHiFi -
16KIS1592K - Forschung Agil and IonLinQ - 13N17230 - Quantum Futur III. The project is co-financed
by ERDF of the European Union and by ‘Fonds of the
Hamburg Ministry of Science, Research, Equalities and
Districts (BWFGB)’.
\end{acknowledgements}

\bibliography{short_version_bibtex.bib}

\appendix

\section{Classical speed limits}
\label{ap:classic}
An $n$ bit key has $2^n$ possible classical states. Without any prior information, the most efficient classical method to find the key of an arbitrary cipher is to guess a key, and test if it can decode an intercepted cipher. When taking $G$ different guesses, the probability that one of them was the correct key is $P_s=G/2^n$. The machine requires $n$ bits, representing the guessed key values and an $n$ bit test-sequence, which is required to check if the guessed key is correct. Hence, $2n$ bits need to be initialized. 
When using a counter to guess the key, on average 2 bits need to be flipped per iteration, meaning on average 1 bit needs to be reset. Hence, to guess an $n$ bit key with probability $P_s$, in total $G+2n$ bits need to be irreveribly reset, each requiring at least the Landauer limit $E_L$ as work input.

Furthermore, if the machine is supposed to finish within a finite time $t$, it must on average be in a new, orthogonal state (testing a new number) every $\delta t=t/G$. Thus, the energy required to start the machine is at least 
\begin{equation}
    E_c\geq G \left(E_L+\frac{h}{4t}\right)+ 2nE_L,
\end{equation}
which simplifies to \eqref{eq:classical_limit}. The contribution to the above energy bound due to the quantum speed limit is not used to counteract entropy, as is the case for bit initialization, and hence may still be available within the system as work after the computation. Thus it could eventually be recuperated, but is necessary to start the machine. Unequal computational speed would potentially lower the energy required to start (if a lower speed is used in the beginning), but eventually more than \eqref{eq:classical_limit} would be required once the search speed is increased to stay within the time limit $t$. Hence a machine operating at constant speed provides the lowest bound. 

Finally, we discuss our definition of initialization. To illustrate the issue we consider a high energy xray photon mode of frequency $f_x$ as a hypothetical qubit. This qubit is with a very high probability in its ground state, and thus could be considered \textit{initialized}. However, this also means, that we cannot perform computation on it, unless we provide work $W\geq hf_x$ to the system. This motivates the definition of the work required to initialize a computational bit as the sum of the work required for initialization into an arbitrary state, i.e.\  $W_\text{init}\geq hf_x$ in the case above. We can easily see that this work exceeds the Landauer limit, as high fidelity initialization requires $hf_x\gg kT \ln 2$. This definition, however, does not permit for a simple, universal time constraint: In the example of classical exhaustive search, the output qubits are set via the computation, not prior to it. Hence, while the $n$ bits representing the cipher text could be limited by the Margoulous Levitin theorem, respectively the speed limit derived in this paper, no additional work is required for the bits of the solution space. Consequently, a case by case analysis is required to account for this effect, also taking into account that the work bound during initialization might be reused during the computation. As setting a memory to a computationally relevant state requires some form of computation, the work requirement due to any quantum speed limit for the computation exceeds that for initialization. Hence we do not account for a speed limit of the initialization separately here.

The result of the search, on a classical machine, is stored in a classical bit, that can be used for further processing. Hence, we do not account for the initialization of readout bits separately for the classical search.


\section{Energy uncertainty of a battery}

We consider a system, that contains Helmholtz free energy, which it can provide to a quantum computer. We call this a battery, albeit it could equally provide the energy mediated by means other than electrochemical reactions. 
Assuming the battery is a closed system that once was in contact with a thermal bath (i.e.\ equivalent to a canonical ensemble) and has $N_\text{DOF}$ accessible degrees of freedom. 
The energy fluctuations in a canonical ensemble are given by 
$$
\Delta E = \sqrt{k_BT^2\frac{\partial \langle E\rangle}{\partial T}}.
$$
When scaling the system, we find $\langle E\rangle\propto N_\text{DOF}$, and hence 
$$
\frac{\Delta E}{\langle E\rangle}\propto \frac{1}{\sqrt{N_\text{DOF}}}.
$$

\section{Limit of gate-based implementation of Grover's algorithm}
\label{ap:gates}

The original, gate-based Grover algorithm requires an oracle ($\hat{O}$) and a diffusion step ($\hat{D}$) per iteration. Agnostic of the platform, implementation of at least one of these operators requires a classical drive signal to create the gates. 
When it is supposed to finish (deterministically find the result) within time $t_F$, these two operations require $\delta t=t_F/(2^{n/2}\pi/4-\frac 1 2)$ on average. Hence, for a computer running with constant gate speed, the classical control signals driving the gates are periodic with $\delta t$, implying that consists of harmonics of the angular frequency $2\pi/\delta t$. Hence, the system requires at least $W_G\geq h/\delta t=h 2^{n/2}\pi/(4t_F)$ as work input to generate the fundamental frequency of the control signal. 
This can be expanded to allow for probabilistic search: The probability of the state  solution evolves with 
$$
P_s(t)=\sin^2\left(\frac{\pi t}{2t_F}-\frac{1}{2^{n/2}}\right).
$$
Thus, we can express the minimum work required to find the solution with probability $P_s$ in time $t$ as 
\begin{equation}\label{eq:si:gatelimit}
    W_Gt\geq\hbar\left(\sqrt{P_s 2^n}-1\right) \left(\pi-2^{1-n/2}\right),
\end{equation}
neglecting initialization and readout, which each require at least $nE_L$ work input.
Naturally, most quantum oracles will be substantially more complex to implement, requiring control signals with higher harmonics.

Similar to the classical search, a computer which runs at a varying clock frequency may start with lower energy, but will require higher frequencies later on to finish within the required time, increasing the lower bound of work required to run the algorithm. 

Note that error correction is not reversible, such that error-corrected codes are subject to Landauer's principle. If $K$ errors are corrected during the run-time, at least one bit needs to be reset each, such that $K E_L$ work input is required additionally. 
Thus overall, a gate-based system requires 
\begin{equation}
    W_\text{G,EC}\geq (2n+K) E_L + W_G
\end{equation}
as work input. While error correction will be necessary for near-term quantum computers, there is no fundamental lower limit to the error rate $K/t$. Hence, for the purpose of finding a fundamental limit, we set $K=0$.

\section{Limit on adiabatic implementation of Grover's algorithm}
\label{ap:adiabatic}

For a conventional adiabatic computer, the Hamiltonian can be described as 
\begin{equation}
    H = (1-c) H_I + (c) H_S + H_C
\end{equation}
where $H_I$ has some easily prepared ground state, and $H_S$ has the solution to the search problem as ground state. The control parameter $c(t)$ is slowly swept from 0 to 1, which is controlled by the Hamiltonian $H_c$. Usually, the trajectory for $c(t)$ is optimized to yield a reliable result as fast as possible. However, this typically involves a tangent-like parameter change \cite{roland_quantum_2002}, with fast leading and trailing edges, corresponding to a rapid change in $c$ when the eigenstates are strongly separated in energy. This corresponds to diverging Fourier components, requiring infinite work to generate. 

We will see below, why adiabatic protocols can intrinsically not saturate the fundamental speed limit. There are two intuitive ways to see that they follow a similar scaling: (1) one can see that the optimized trajectories describe a pulse with width $\delta t \sim t_F/2^{n/2}$ at the beginning and end of the algorithm, requiring $\mathcal{O}(\hbar 2^{n/2}/t_F)$ work input to generate. (2) Even though the state of an adiabatic computer is close to the ground state, a finite speed of parameter change induces a minuscule contribution of higher lying eigenstates to the wavefunction. Hence, the overall work must be at least as high as the energy of the second lowest eigenstate, i.e.\ $\mathcal{O}(\hbar 2^{n/2}/t_F)$.

Finally, we note that a shortcut to adiabaticity is not possible under these conditions, as counterdiabatic driving effectively requires knowledge of the solution.

\section{Active control}
\label{ap:eq45}
When the search task at hand is truely random, i.e\ there is no prior to any specific solution state, the algorithm should also not exhibit any preference. Hence, the initial state must be an equal superposition of all possible solutions. Without loss of generality, we choose equal phases for all states in the solution space $k\in{0,1}^n$, such that the initial state is $\left|i\right\rangle=\sum_k \left|k\right\rangle/2^{n/2}$ for the computational space of $H_Q$. 

Equally, for a random search problem, there is no information present such as energy gradients, other than the possibility to confirm a specific solution $\left|s\right\rangle$. Hence, without bias, the Hamiltonian cannot include any information other than the initial and the solution state, As well as their respective energies. 

Consequently, the Hamiltonian must be of the form \eqref{eq:gen_ham}. This allows for computing the change in probability $P_s=|\langle s|\psi\rangle|^2$ of the computational state overlapping with the solution
\begin{align}\label{eq:ap:dpdt}
    \frac{dP_s}{dt} &= 2 \text{Im}\left(\langle \psi|s\rangle\langle s|H|\psi\rangle\right)\nonumber\\
    &= \frac{2\omega_i}{2^{n/2}} \text{Im}\left(\langle \psi|s\rangle \langle i|\psi\rangle\right)\\
    &=2\frac{\omega+\delta\omega}{2^{n/2}} \sin(\alpha_{ab})|\langle s|\psi\rangle| |\langle i|\psi\rangle|\nonumber
\end{align}
with the angle $\alpha_{ab}=\arg(\langle i| \psi\rangle\langle \psi| s\rangle)$, the average frequency $\omega=(\omega_i+\omega_s)/2$, and the detuning $\delta\omega=(\omega_i-\omega_s)/2$. This can equivalently be written as \eqref{eq:rateofchange}.
We note that all eigenstates of \eqref{eq:gen_ham} have real-valued relative amplitudes, such that $\text{Im}\left(\langle \psi|s\rangle \langle i|\psi\rangle\right)=0$. Hence, adiabatic protocols, which remain close to an eigenstate, have small $\alpha_{ab}$, and thus progress relatively slowly. 
As $\left|s\right\rangle$ and $\left|i\right\rangle$ are nearly orthogonal, the angle $\alpha_{ab}$ is similar to the (squared) Bures, or Fubini-Study angle. This allows for a geometric interpretation of the state on the Bloch sphere of the two qubit space spanned by $\left|s\right\rangle$ and $\left|i\right\rangle$. In contrast to quantum speed limits, we consider the angle between the state and the solution state, rather than with itself at different times \cite{wu_quantum_2020}. 
To maximize knowledge gain, i.e.\ the rate of change $\frac{dP_s}{dt}$, an algorithm needs to maximize $\sin(\alpha_{ab})$, respectively $\omega$ and $\delta \omega$, which may not be achieved simultaneously, due to the dynamics of $\alpha_{ab}$. 

To investigate this, we evaluate the axis around which the state $|\psi\rangle$ rotates. When the rotation axis is closely aligned with $\left|s\right\rangle$ and $\left|i\right\rangle$, the sign of $\alpha_{ab}$ will rapidly change, reducing the overall effect. To investigate the dynamics of $\alpha_{ab}$, we compute the time derivative
of $A=\langle \psi|s\rangle \langle i|\psi\rangle$
\begin{equation}
    \frac{d}{dt} A = -i \delta \omega \left(A-\frac{P_i+P_s}{2^{n/2}}\right) + i\frac{\omega}{2^{n/2}}(P_i-P_s), \label{eq:A_dyn}
\end{equation}
with the probability $P_i=|\langle i|\psi\rangle|^2$. 
As $P_i+P_s\approx 1$, we find $A-2^{-n/2}$ rotating rapidly with detuning $\delta\omega$ due to the first term, with small correction $\order(\delta \omega 2^{-3n/2})$. When $|A|$, $\delta\omega/\omega \gg 2^{-n/2}$ and constant, we find that amplitude of $A$ is changing slowly. In this case, the Ansatz $A=\bar{A}e^{-i\alpha_{a,b}(t)}$ yields 
$$
\frac{d}{dt}\alpha_{a,b}(t)\approx \delta \omega.
$$

Substantial information gain over longer times,  i.e.\ $\langle\text{Im}(A)\rangle_{\Delta t}\gg0$, where $\langle\cdot \rangle_{\Delta t}>0$ denotes the average over a finite time interval $t_1\rightarrow t_1+\Delta t$ during the computational time $t$, requires $\langle\delta\omega\rangle_{\Delta t}$ to be small with respect to the inverse of the computational time $\Delta t$, as otherwise $\frac{dP_s}{dt}$ changes sign. This can be achieved by $\delta\omega\sim0$ or the sign of $\delta \omega$ can be demodulated, using the control Hamiltonian $H_C$, albeit the latter requiring work input for active control of the system. 
In the following, we will first investigate the effects of constant detuning, followed by modulated detuning, to derive limits on the information gain. 

To limit the possible detunings, we derive the average information gain over a time $\Delta t$, which is substantially shorter than the overall computing time, such that $P_i$ and $P_s$ only change insignificantly, but much longer than the time scale $1/\omega$. Integrating \eqref{eq:A_dyn}, we find 
\begin{align}\label{eq:A_time_average}
    \langle A\rangle_{\Delta t}\!=&A_0 e^{-i\frac{\delta \omega \Delta t}{2}} \text{sinc}\left(\frac{\delta \omega \Delta t}{2}\right) +\\
    &\frac{1+\frac{\omega \langle P_i\!-\!P_s\rangle_{\Delta t}}{\delta \omega }}{\sqrt{2^n}} \left(1-e^{-i\frac{\delta \omega \Delta t}{2}}\text{sinc}\!\left(\!\frac{\delta \omega \Delta t}{2}\!\right)\right)\nonumber
\end{align}
with the unnormalized sinc function $\text{sinc}(x)=\sin(x)/x$, and the initial value $A_0=A(t_1)$. 
The first term renders large detunings $\delta\omega>2\pi/\Delta t$ strongly unfavorable, decreasing quadratically in $\delta\omega\Delta t$. The second term, albeit small, linearly increases with $\delta\omega$, hinting at a small, but non-zero optimal detuning. To find a limit for the optimal detunig $\delta\omega_o$, we locally optimize the time-averaged information gain $\frac{dP_s}{dt}$ with respect to the detuning. As we know $\delta\omega_o$ is small ($\frac{\delta \omega_o\Delta t}{2}\ll1$) due to the $\text{sinc}$-term we expand \eqref{eq:A_time_average} to second order in $\delta\omega$, finding the approximation
\begin{equation}
    \frac{\delta \omega_o\Delta t}{2}\!\approx\! \frac 3 C \frac{\text{Im}(A_0)+\frac{C }{ 2^{n/2}}\langle P_i\!-\!P_s\rangle_{\Delta t}-C\text{Re}\!\left(A_0\!-\!\frac{1}{2^{\frac n 2}}\right)}{ \text{Im}(A_0)+\frac{C }{ 2^{n/2}}\langle P_i\!-\!P_s\rangle_{\Delta t} +\frac 6 C \text{Re}\!\left(A_0\!-\!\frac{1}{2^{\frac n 2}}\right)}, 
\end{equation}
with a time constant $C=\frac{\omega\Delta t}{2}\gg1$, yet small enough to not substantially effect the probabilities $P_i$ and $P_s$, $C\ll2^{n/2}$. We consider two relevant scenarios in particular: the beginning and end of the algorithm, as well as the middle part. 

(1) Immediately at the beginning, $|\psi\rangle=|i\rangle$, such that $\text{Im}(A_0)=0$ and $\text{Re}(A_0)=2^{-n/2}$. This yields $\frac{\delta \omega_o\Delta t}{2}\approx3/C$, which maximizes the growth of the imaginary component of $A$, rotating the real-valued part to $\alpha_{ab}\approx\pi/2$, i.e.\ $A$ is almost entirely imaginary, and thus optimal for the information gain.
Conversely, the same holds when $|\psi\rangle$ approaches $|s\rangle$, yielding $\frac{\delta \omega_o\Delta t}{2}\approx3/C$ within the final $\Delta t$ of execution time before the solution is reached. 

(2) During the majority of the run time of the algorithm, the information gain should be optimal, meaning that $A$ is almost entirely imaginary. We can thus approximate $\text{Re}(A_0)\ll\text{Im}(A_0)$, finding 
\begin{equation}\label{eq:ap:det_runtime}
\frac{\delta \omega_o\Delta t}{2}\approx \frac 3 C -  \frac{\text{Re}\left(A_0-\frac{1}{2^{\frac n 2}}\right)}{\sqrt{P_iP_s}+C\frac{\langle P_i\!-\!P_s\rangle_{\Delta t}}{2^{n/2}}}.    
\end{equation}
This holds for most parts of the algorithm except when the state is close to $|i\rangle$ or $|s\rangle$, when the real part becomes relevant, see above. 
This means, that the approximation \eqref{eq:ap:det_runtime} requires 
$P_i\geq 4C^2/2^{n}$, 
such that the denominator does not diverge. 
Immediately before finding the solution, the the optimal detuning allows for rotating the state $|\psi\rangle$ to overlap exactly with $|s\rangle$, and thus can strongly depend on the prior protocol. Here we require the algorithm to locally operate with high knowledge gain, i.e.\ $\alpha_{ab} \approx \pi/2$ before the final approach of the solution $|s\rangle$. In this case, we can limit the optimal detuning to 
\begin{equation}\label{eq:ap:optimal_detuning}
    \frac{\delta \omega_o\Delta t}{2}\leq \frac 4 C.
\end{equation}

Hence, an algorithm that varies the detuning $\delta \omega$ slowly with a characteristic time $\Delta t$, can maximally achieve a knowledge gain limited by 
\begin{equation}\label{eq:si:relative_rate}
    \frac{dP_s}{dt}\leq 2\frac{\omega+\frac{16}{\omega\Delta t^2}}{2^{n/2}}|\langle s|\psi\rangle\langle i|\psi\rangle|.
\end{equation}
As the required control bandwidth, we conservatively estimate the half width half maximum of a moving average of $\Delta t$, resulting in $\omega_\text{hwhm}\geq 3.79/\Delta t$. In order to attribute for time-scales comparable to the entire computational time $t_F$ it takes to complete the algorithm, i.e.\ find the solution deterministically, we further introduce a corresponding frequency cut-off $\omega_\text{cut}=3.79/t_F$, such that we estimate the control bandwidth $\omega_\text{ctrl}=\text{max}(\omega_\text{hwhm}-\omega_\text{cut}, 0)$. Alternatively, when the algorithm is only suppost to generate a probabilistic result, one can also set $\omega_\text{cut}=3.79/t$. This cut-off frequency allows for considering very slow control sequences without diving into the details of the continuity of control sequences before and after the actual quantum search, i.e.\ during initialization and read out.  

Together with the eigenenergies of \eqref{eq:gen_ham}
\begin{equation}\label{eq:si:eigeneergies}
    E_{e,\pm}=\hbar\omega\pm\hbar \sqrt{\delta \omega^2+(\omega^2-\delta \omega^2)/2^n }, 
\end{equation}
we can write the required thermodynamic work as $W\geq E_{e,+} + \hbar\omega_c+W_\text{init}$, with the work required for initialization and readout $W_\text{init}\geq2nE_L$. 

We can thus bound the average frequency 
\begin{equation}\label{eq:si:limit_omega}
    \omega\leq \frac{\tilde{W}}{\hbar (1+2^{-n/2})}-  \omega_\text{ctrl}
\end{equation}
with the work $\tilde{W}=W-W_\text{init}$ input excluding initialization and readout, which is tight for $\delta \omega \approx 0$. 
Hence, the knowledge gain is limited by 
\begin{equation}\label{eq:ap:gain_limitC}
    \frac{dP_s}{dt}\leq 
2\frac{\left(\frac{\tilde{W}}{\hbar (1+2^{-n/2})}-  \omega_\text{ctrl}\right)\left(1+\frac{4}{C^2}\right)}{2^{n/2}}|\langle s|\psi\rangle\langle i|\psi\rangle|.
\end{equation}
For $\Delta t \ll T$, $1/C^2\propto \omega_\text{ctrl}^2$, and hence the expression is maximized when no active control is performed, i.e. for $\omega_\text{ctrl}=0$. In this case, \eqref{eq:ap:gain_limitC} simplifies to \eqref{eq:rateofchange3} with $b=\frac{1+(\frac{4}{\omega t})^2}{1+2^{-n/2}}$, assuming $\alpha_{ab}\approx\pi/2$.

Finally, we need to confirm that demodulation of larger detunings is not efficient.  We investigate a modulated detuning $\delta \omega(t)=\delta\omega_o \text{sin}(\omega_c t)$. For the control frequency ratio $r=|\delta \omega_o/\omega_c|\ll1$, the time averaged amplitude $\langle A\rangle_{\Delta t}\approx  A_0 (1-r^2/4)$. Hence, higher modulations increase the work input requirement, as well as decrease the knowledge gain rate, such that we further can exclude fast active control. 

In summary, neither fast nor slow control can provide an advantage in information gain rate that outweighs the opportunity cost of the work requirements associated with active control of the system.

\section{Fundamental limit}
\label{ap:fundamental_limit}
To establish a fundamental limit, we need to further restrict the range of detunings that provide optimal information gain: In the previous section, we found that without active control, the optimal detuning is limited to $\delta\omega_o\leq \frac{16}{\omega t^2}$. This provides a strict limit for long integration times, but can be relatively large for short times.
Hence, in the initial phase of the algorithm, rather than searching the detuning which maximized information gain \eqref{eq:ap:optimal_detuning}, we need to optimize it for a fixed work input.

First, we linearize \eqref{eq:A_time_average} in time, yielding $\text{Im}\langle A\rangle_{t} t\approx \frac{\omega}{2^{n/2}}\frac{t}{2}$, for $t_1=0$ and $\Delta t=t\ll 2/\delta \omega$. Inserting this in \eqref{eq:ap:dpdt}, and integrating, we obtain the estimate 
\begin{equation}
\label{eq:ap:short_times}
    P_s(t)\leq \frac{ (1+k) \omega^2 t^2}{2^n} +P_s(0),
\end{equation}
with the relative detuning $k=\delta\omega/\omega$. 
To better understand the implications of a fixed work input, we first assume a relatively large detuning $k\gg 2^{-\frac n 2}$, such that the eigenenergy \eqref{eq:si:eigeneergies} is approximately $E_{e,+}\approx \hbar \omega (1+k)$. 
Using \eqref{eq:ap:short_times}, 
\begin{equation}
    P_s(t)\leq \frac{ (1+k) t^2}{2^n}  \frac{\tilde W^2}{\hbar (1+k)^2}+P_s(0),
\end{equation}
which is maximized for $k=0$, which is out of bounds for this approximation, i.e\ the best detuning must be small. 

To find the actual value of the best detuning, we repeat this process with the full expression for the eigenvalues \eqref{eq:si:eigeneergies} and maximize the bound \eqref{eq:ap:short_times}. 
This yields  
\begin{equation}\label{eq:ap:limitPs}
    P_s(t)\leq b\frac{\tilde W^2 t^2}{\hbar^2 2^n}  +P_s(0),
\end{equation}
with the prefactor $b=\frac{1+k}{(1+\sqrt{k^2+2^{-n}(1-k^2) })^2}$. This is maximal for $k\omega = \delta\omega_o \approx \frac{\omega}{\sqrt{3\cdot 2^{n}} }$, bounding the prefactor to 
$b\leq 1-\frac{1}{\sqrt{3\cdot 2^{n}}}+\order(2^{-\frac 3 4 n})< 1$.

For sufficiently large $n$, 
$$\frac{4}{\omega}\left(3\cdot 2^{n}\right)^{\frac 1 4}\leq t \ll \frac 2 {\delta \omega_o}\approx \frac{\sqrt{12\cdot 2^{n}} }{\omega},$$ 
such that $\frac{16}{\omega^2 t^2}\leq \frac{1}{\sqrt{3\cdot 2^{n}}}$. Hence, we find that $b\leq 1$ for all times, consistent with \eqref{eq:ap:gain_limitC}.

Hence, we can rewrite \eqref{eq:ap:limitPs} as
\begin{equation}\label{eq:ap:limit}
    \tilde{W}\geq \sqrt{P_s 2^{n}-1}\frac{\hbar}{t}.
\end{equation}
With the approximate lower limit of required work defined as $W_0t\equiv\hbar \sqrt{2^nP_s(t)-1}$, we find that the gate-based Grover algorithm requires approximately $\pi$ times the work input, $W_G\approx \pi W_0$.

\section{Initialization and readout}
\label{ap:init}
The limits \eqref{eq:ap:limit} and \eqref{eq:si:gatelimit} contain an offset, such that for $P_s=2^{-n/2}$ no work input is required: $\tilde{W}=W_G=0$. Of course, this does not indicate a possible work around, but is due to neglecting the initialization and readout process. In both cases, an array of $n$ bits need to be initialized: The computer itself requires at least $n$ qubits to represent the states, and the readout system requires at least $n$ classical bits to copy the result to. When the solution to a symmetrically encrypted cipher text is searched, additionally a sufficient number of bits to represent the cipher text, and a condition indicating a successful decryption are required, each requiring $n$ bits as well. Hence, the work required to initialize the computer and read out the result are $W_\text{init}\geq 2nE_L$ for a general search problem, respectively $W_\text{init}\geq 4nE_L$ for a known plain text attack on a symmetrically encrypted data channel. 

\section{BHT}
\label{ap:bht}
The analysis can naturally be extended for algorithms with a relevant memory space requirement, such as the Brassard-Hoyer-Tapp collision search (BHT) \cite{BHT1998}, which we sketch here: For a two-to-one function $F:X\rightarrow K$ with n-bit image space $K$, BHT used $k$ classical evaluations of F, stores the output in memory, and then performs a Grover search over the rest of the domain. 
Using an effective Hamiltonian $H_F = \hbar \omega \ket{+_\kappa}\bra{+_\kappa}$, representing the function $F(\kappa) = x_K$ with $\ket{+_K}=(\ket{0^n}+\ket{x_K})/\sqrt 2$, $\kappa\in K$, $x_\kappa\in X$, we find the minimum work requirement for a single evaluation within $\delta t$ is $W_\text{single}\geq h/4\delta t$ according to \eqref{eq:work}, respectively the Margolous Levitin theorem. 

Hence, we find for the minimum work requirement
$$
W_\text{BHT}\geq k (n+1) E_L + \max{\left( \frac{h}{4\delta t} , \sqrt{\frac{2^{n}}{k}P_s-1}\frac{\hbar}{t_s}\right)}
$$
with the total time of the algorithm $t_T=k\delta t+t_s$, considering the work stored in the computer during initial evaluation of the functions can be used again to power the quantum computer.  





Hence, we can optimize the time split, such that both parts of the algorithm match in their speed limit requirement, effectively minimizing the overall work requirement. This results in the optimal time spent on the quantum search
$$
t_s=\frac{t_T}{k \frac{2\pi}{4 \sqrt{\frac{2^{n}}{k}P_s-1}} +1}. 
$$
This results in the work requirement, which only depends on the number of samples $k$:





$$
W_\text{BHT}\geq k (n+1) E_L + k \frac{h}{4t_T} + \sqrt{\frac{2^{n}}{k}P_s-1}\frac{ \hbar }{  t_T}
$$
For the optimal number of samples that minimizes the work requirement, we find




$$
k =   \frac{2^{n/3} P_s^{1/3}}{\left((n+1) E_L  \frac{4t_T}{\hbar} + 2\pi\right)^{2/3}}.
$$
As $E_L\gg \hbar/t_T$ the optimum will be reached for $k\ll 2^{n/3}P_s^{1/3}$. Thus the fundamental thermodynamic limit for collision search, only depending on available time $t_T$, bath temperature $T$ and allowed success probability $P_s$ is 



$$
W_\text{BHT}\geq 2^{n/3} P_s^{1/3}\left((n+1) E_L  \frac{4t_T}{\hbar} + 2\pi\right)^{1/3}  \frac{5}{4}\frac{ \hbar }{  t_T}.
$$

We can use the same conditions as in table \ref{tab:quantum_key_lengths} to find the hash lengths required for a comparable security level. For example, for 128 bit of classical security, corresponding to 256 bit hash functions in the classical security framework, or NIST PQC security level 2, we find under the conditions of line 1 in table  \ref{tab:quantum_key_lengths}, that the hash function needs an image space of at least 415 bit. 

For 256 bit of security, corresponding to a classical 512 bit hash function, requires 788 bit image space under the condition in line 2. The cosmic limit in line 3 corresponds to a hash function with 1077 bit image space. We note that the bit number for hash functions less than doubles, consistent with the naive interpretation of the BHT scaling, predicting a 50\% increase in image space dimension for quantum resistance. This differs from the results for Grover's algorithm in the main text, as an increase in time does not benefit the BHT performance as much. It will effectively still be limited by the classical sampling before the quantum search part of the algorithm.

\section{Ballistic quantum search tightness}
\label{ap:ballistic}
Here, the computer is initialized to $|i\rangle$, and $\delta\omega=0$. As the eigenstates of the Hamiltonian \eqref{eq:gen_ham} are aligned with $|i\rangle\pm|s\rangle$, and the energy splitting is $\Delta E_e=2\hbar\omega/2^{n/2}$
\begin{equation}\label{eq:ap:prob_ballistic}
    P_S(t) = \frac{1}{2^n} + \left(1-\frac{1}{2^n}\right) \sin^2\left(\frac{\omega}{\sqrt{2^n}}t\right).
\end{equation}
The solution is found deterministically after $t_F=\frac{\pi\sqrt{2^n}}{2\omega}$. Following \eqref{eq:si:limit_omega}, we find 
$\hbar\omega\leq \tilde{W}/(1+2^{-n/2})$, and as \eqref{eq:ap:prob_ballistic} is monotonically increasing from $t=0$ to $t_F$, we can bound the probability for finding the solution as \eqref{eq:ballistic}. 
In particular, we can conclude that the solution cannot be found deterministically faster than 
\begin{equation}
    t_F\geq \frac \pi 2 \frac{\sqrt{2^n}+1}{\tilde{W}}
\end{equation}
For short term information gain, we use \eqref{eq:A_dyn}, setting $\langle i\mid\psi\rangle\approx1$, and thus find
$$
\langle s\mid\psi\rangle \approx i\frac{\omega t}{\sqrt{2^n}}+\frac{1}{\sqrt{2^n}},
$$
which provides an estimate for the probability of finding the solution 
$$
\sqrt{P_s(t)2^{n}-1} \approx \omega t.
$$
As $|\langle i\mid\psi\rangle|$  decreases as $P_s$ increases, we find can bound the work required by the system is bound by
$$
\tilde W\geq \sqrt{P_s(t)2^{n}-1}\frac \hbar t \cdot (1+2^{-n/2})=W_B,
$$
The bound $W_B$ approaches the optimal prefactor $b$ \eqref{eq:ap:limitPs} to a factor better than $W_B/W_0\sim 1+\frac{1}{\sqrt{2^{n}}}$, i.e.\ it is asymptotically tight.

\section{Cosmic key length limits}
\label{ap:key_lengths}
While various energy and time scales could be considered reasonable, here we aim to investigate fundamental limitations. According to cosmological models, in correspondence with astronomical observations, the universe is entering a phase dominated by a cosmological constant, corresponding to an accelerating expansion. This causes a distance $x_0$ at time $t=0$ to increase $x(t)=a(t)x_0$, according to the scaling factor $a$. This means, that signal we send out now, cannot reach an observer that is further away than $c/\frac{da}{dt}$: When sending out a photon towards a very distant galaxy, the photon will never reach it, as the space in between is expanding faster than the speed of light $c$. As there is a finite energy density within this cosmological event horizon, this means we can fundamentally only access a finite amount of work. To estimate this upper bound on possibly available energy, we approximate the mass of matter (bright and dark matter) within the cosmological event horizon. As we are primarily interested in an upper limit, rather than a precise value, we assume a simplified model of a Friedmann–Lemaître–Robertson–Walker metric.
At time $t=0$, all matter within the event horizon is sent to an adversary, converted to energy,
which is used to run a quantum search on a local quantum computer. The system is assumed small enough to withstand the expansion of the universe, but large enough to avoid collapse to a black hole. 

Of course, stricter models could be constructed, considering that the adversary would need to collect all the matter, i.e.\ send out a coordination signal, and that transport of all matter or energy to one location has some inefficiencies due to fundamental physical laws. These effects, however, only reduce the possibly available work by an order of magnitude, and thus only modify the number of bits by less than 1\%. 
During the expansion of the universe, the matter density will decrease, and thus the Hubble constant will decrease. To find an upper limit of available energy, we assume the (baryonic and dark) matter density observed today, but the minimal Hubble constant of a late-stage, dark energy dominated universe, corresponding to a scale factor $a(t)=\exp{H_0\sqrt{\Omega_\Lambda}t}$, with todays Hubble constant $H_0$ and the dark energy density $\Omega_\Lambda$.
Further, we ignore speed limits due to finite propagation speed of information within the volume of the computer. As few qubits are required to implement the algorithm, the Bremermann–Bekenstein limit is not relevant, but rather the physical energy density, which dilates time within the computer, i.e.\ when its density becomes too high, its operation speed as seen by a distant observer is reduced. We obtain an upper bound to possible search space, by assuming the density is sufficiently low, such that time dilation is irrelevant, but its size is small enough for propagation delay to be negligible.

The maximum energy available in this scenario is 
\begin{align*}
    W_\textrm{cosmic}&\leq \frac{4}{3}\pi \left(\frac{c}{\sqrt{\Omega_\Lambda} H_0}\right)^3\rho_m c^2\\
    &=\frac{(1-\Omega_\Lambda) c^5}{2H_0{\Omega_\Lambda}^\frac{3}{2}G}\\
    &=4.62\cdot 10^{69} \si{\joule}
\end{align*}
using the cosmological parameters of the Planck collaboration \cite{aghanim2020planck}. 
(Hubble constant $H_0\approx 67.36 \frac{\si{\km}}{\text{Mpc}\si{\second}}$, matter density $\rho_m \approx 2.69\cdot 10^{-27}$~kg/m$^3$, normalized dark energy density $\Omega_\Lambda\approx 0.6847$)

Providing time until star formation is expected to cease, $t=10^{14}$ years, 
we find that an ideal ballistic quantum search would be able to reconstruct a $830$~bit key deterministically. 
Hence, we can be confident that a key, which is substantially longer than $n\gg 830$ bits, cannot be reconstructed by exhaustive search, independent of the technological advancement of any adversary, based on thermodynamics, fundamental physics, and cosmological observations.

When requiring a higher security level, i.e.\ a smaller probability, for the key to be recovered, we can use \eqref{eq:qlimit}. For same conditions as described above, we find the largest key that can be recovered with a probability of at least $P_s \geq 10^{-12}$, is $871$~bit long.
Hence, an unbroken symmetric cipher with a key longer than $872$~bit is effectively as secure as a one time pad encryption up to a probability of $P_s \geq 10^{-12}$ from a thermodynamic perspective.

\end{document}